# Electron Multi-Beams ("Electron Grids", "Electron Lattices" and "Electron Grates"): New Elements for Accelerators

*Vladimir Shiltsev*

*Fermi National Accelerator Laboratory, PO Box 500, Batavia, IL 60510, USA*

*Abstract*

Since the initial proposal in 1990's, the method of electron lenses has been successfully developed and employed at the high energy particle colliders, like Tevatron and RHIC. Here we propose a new set of electron multi-beam elements – electron grids, electron lattices, electron grates and other – which employ many beams in various spatial configurations, with possible time-varying currents. We present major principles of operation and generation of such systems and briefly discuss a set of possible applications in accelerators.

## 1. Introduction

The electron lenses were initially proposed for beam-beam compensation in superconducting hadron colliders but eventually found many other applications – see detail history account and comprehensive overview in Ref. [1]. In essence, the electron lens is a low-energy, high-current density, magnetically confined electron beam whose electromagnetic fields can be used for active manipulation of high-energy beams in accelerators. For axially symmetric electron current-density distribution $j_e(r)$, the force exerted by electron lens on the high-energy test particle, say, proton, going through it parallel to the lens axis is equal to:

$$F_r(r,t) = e(E_r + \beta_p B_\theta) = (1 \mp \beta_p \beta_e) \frac{4\pi e}{c\beta_e} \left( \frac{1}{r} \int_0^r j_e(r,t) r dr \right) \quad . \tag{1}$$

There "-" sign in the first factor corresponds to electron and test particle velocities $v_{e,p}=c\beta_{e,p}$ being parallel, and "+" to antiparallel case. For example, a 3 Ampere 1 mm diameter beam of 10 kV electrons generates ~1 MV/m fields. The orthogonal configuration of the beams is quite possible as well [2]. The EM force Eq.(1) is linear at the distances smaller than the characteristic beam radius $r<a_e$ but scales as $1/r$ for $r>a_e$.

Four key advantages promote application of electron lenses in high energy accelerators:
a. besides the EM forces, the low-energy electron beam is "transparent" and poses no nuclear interactions, thus can be placed in high energy beams without common, material-related risks of damage;
b. electrons interact with the high-energy particles only once on each turn, leaving no room for excitation of coherent instabilities;
c. immersion of the electron beam in strong longitudinal magnetic fields allows the electron current profile $j_e(r)$ – set in the external electron sources - and thus the EM field profiles to be easily changed for different applications;
d. the electron beam current can be varied quickly on very fast time scales $O(10ns)$.

Correspondingly, the applications of the electron lenses are numerous: i) they have been built, commissioned and made operational for beam-beam compensation in the Tevatron proton-antiproton collider [3, 4, 5, 6] and in the Relativistic Heavy Ion Collider (RHIC) [7, 8, 9]; ii) the Tevatron electron lenses were used during regular collider operations for longitudinal beam collimation – removal of uncaptured protons and antiprotons from the abort gaps [10]; iii) transverse halo collimation by hollow electron beam has been successful demonstrated in the Tevatron in 2010–2011 [11]; iv) electron-lens compensation of space-charge effects in high-intensity proton accelerators, including super-collider injectors, was proposed in 2000 [12] and is now subject of the experimental accelerator research program at the IOTA test ring at Fermilab [13]; v) proper choice of the electron beam current profile $j_e(r)$ was shown to lead to nonlinear integrability in the accelerator and collider focusing lattices [14, 3, 13], vi) electron lenses were also proposed for selective slow extraction system [15], for enhanced Landau damping of coherent instabilities [17]; and for the "beam-beam" kicker operation [2].

Below we propose and discuss yet another set of opportunities which can be offered by a variety of systems with multiple electron beams for particle accelerators.

## 2. Electron Multi-Beam (EMB) systems

To start with, let's consider an array of electron beams regularly spaced along a line – see Fig.1 a). If we take for simplicity, the current distribution in each beamlet to be Guassian $j_e(r) \sim \exp(-r^2/2 a_e^2)$, then the resulting EM field of the array will have the functional form of :

$$\vec{F}(\vec{r}) \sim \sum_m \left(\frac{\vec{r}_m}{r_m}\right) \times \frac{1-e^{-r_m^2/2a_e^2}}{r_m} \qquad (2)$$

where $\vec{r}_m = \vec{r} - m \cdot \vec{s}_o$, see Fig.2 b). Therefore, the EM field configuration is determined by the ratio of the period $s_0$ to the beamlet size $a_e$ - e.g., $s_0 = 20$ and $a_e = 1$ for the distributions depicted in Fig.2. In the case when the high energy beam size $\sigma$ is large enough so several electron beamlets can fit within, one can foresee opportunities to affect different parts of the beam differently.

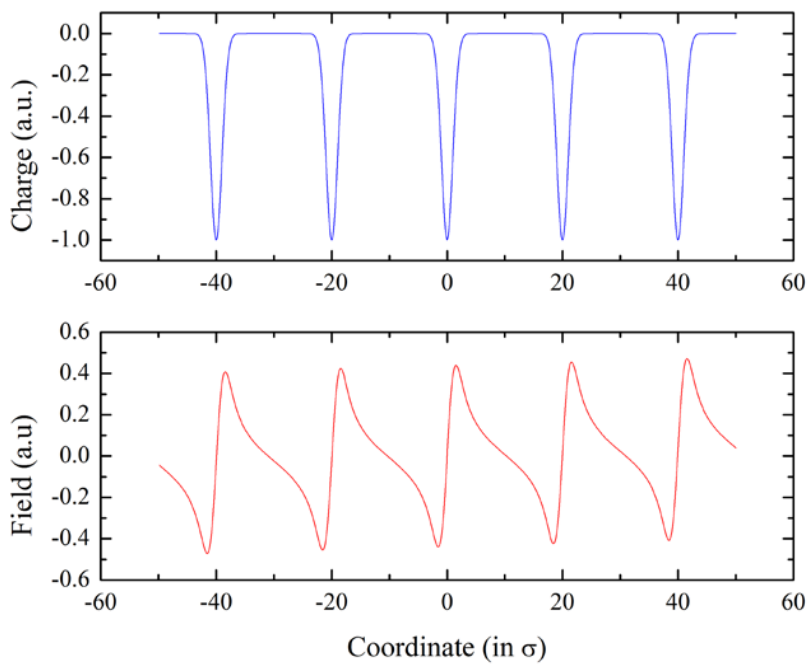

**Figure 1:** a) charge density in the linear array of regularly spaced electron beams with Gaussian current profiles; b) electric field generated by the EMB array depicted above.

A variety of practical configurations can be imagined – from "electron lattice" of 2D regularly spaced beamlets parallel to the high-energy beam direction (z-axis), see Fig.2; to series of parallel beams in an "electron grate" as in Fig.3a or in an "electron grid" as in Fig.3b; to various combinations of axially symmetric electron rings constituting "electron target"-like structures as pictures in Figs. 4a) and b).

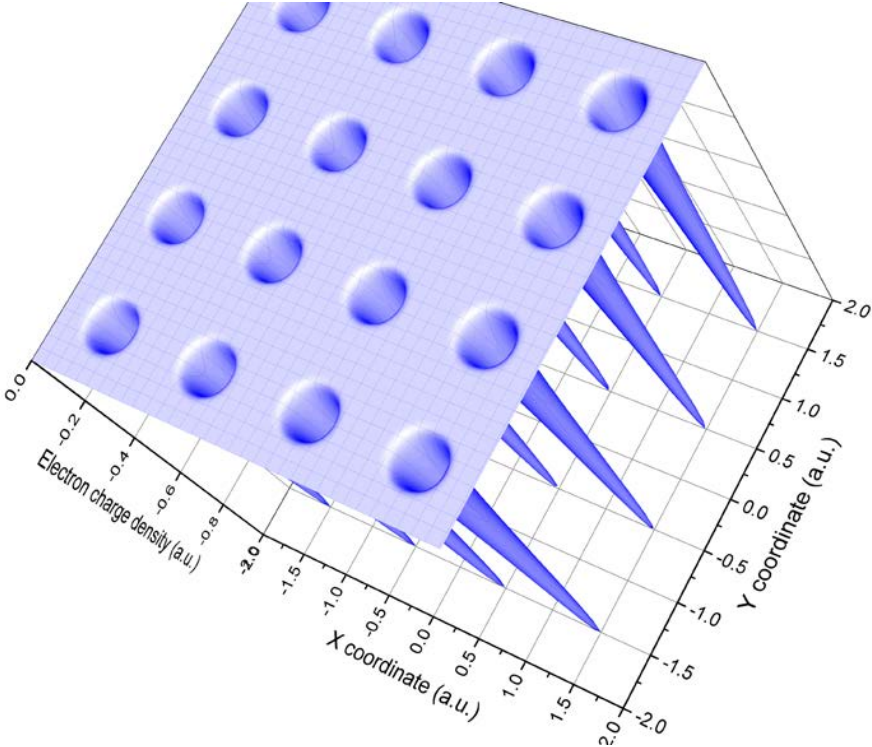

**Figure 2:** "Electron lattice" of 2D regularly spaced Gaussian electron beamlets moving along z-axis.

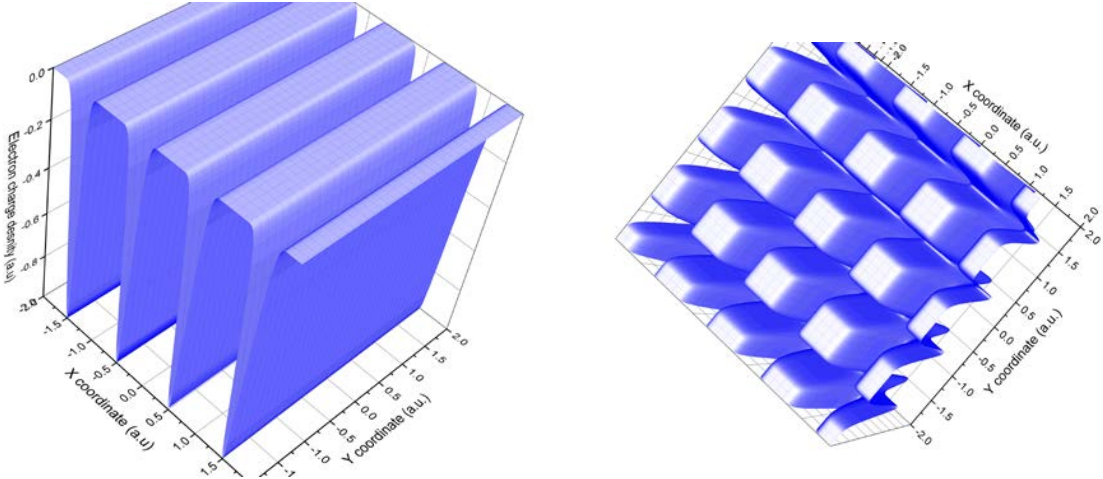

**Figure 3:** Electron multi-beam structures: a) left - "electron grate", and b) "electron grid".

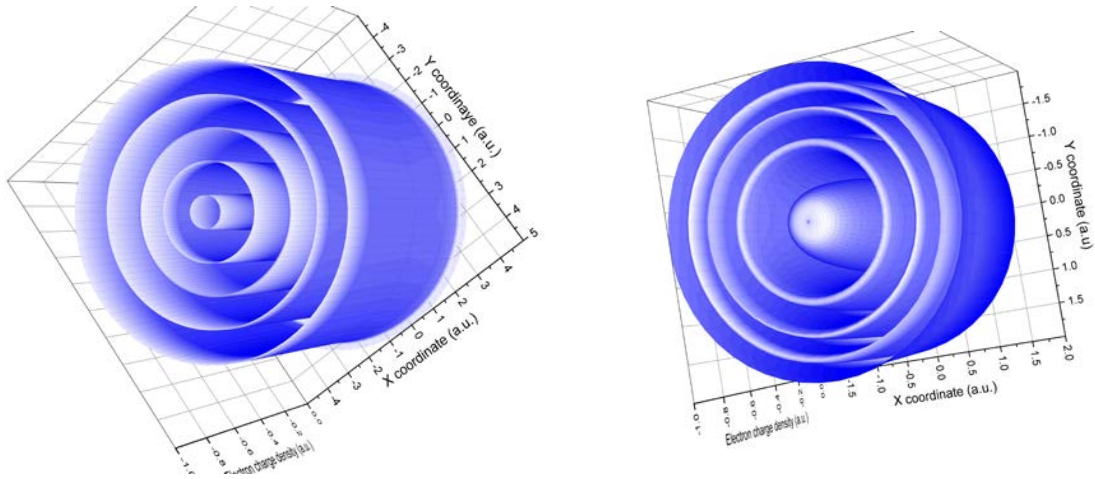

**Figure 4:** Axially symmetric "electron target" multi-beam structures: a) left – with equally spaced electron rings, and b) right – with Fresnel-lens-like electron rings.

## 3. Technical aspects of the EMB generation

The electron multi-beams (EMB) technology can greatly copy that of the electron lenses which was well developed during the construction and upgrades of the above-mentioned systems built for and installed in the Tevatron and RHIC. They employed some 10 kV, Ampere-class electron beams of millimeter to submillimeter sizes with a variety of the transverse current distributions $j_e(r)$ generated at the thermionic electron gun, including flat, Gaussian and hollow ones. The electron beams in the lenses as well beamlets in the EMB can be made very stable transversely being immersed in a strong magnetic field - about $B_g$=0.1-0.3 T at the electron gun cathode and some $B_m$=1.0-6.5 T inside a few meters long main superconducting solenoids. The electron beam transverse alignment on the high-energy beam is done by trajectory correctors to better than 20-50 microns, which is usually a small fraction of the rms high-energy beam size $\sigma$. The latter is usually in the range of a mm to several hundreds of microns, thus, to be able to explore different parts of the high-energy beam, the electron beamlet size $a_e$ should be of the order of ten(s) of microns. That can be done in an electron lens type magnetic system which adiabatically compresses the electron-beam cross-section area in the interaction region by the factor of $B_m/B_g \approx 10$ (usually, variable from 2 to 60), proportionally increasing the current density $j_e$ of the electron beam in the interaction region compared to its value on the gun cathode, usually of about $j_{cath}$ =2-10 A/cm$^2$.

As one can see from Eq.(1) the maximum EM force from a single beamlet scales as ~ $j_e\, a_e$. Thus, if strong forces are required at small beamlet size $a_e$ then one might pursue both highest current density cathodes and the large magnetic compression ratio as $j_e = j_{cath}(B_m/B_g)$. Technologically, the required geometry beamlets can be generated either by employment of individual cathodes for each of them or by appropriate alternation of emitting and non-emitting areas on a single cathode, as it was done, e.g. for generation of a hollow electron beam for proton collimation in the Tevatron [11]. Yet another possibility would be to use a properly shaped near-cathode mask which would stop emission everywhere but at the locations of the holes, but such approach might have limitation due to thermal issues at high currents. It is generally considered that high current electron beam interception should be avoided.

An interesting opportunity deserving further study is offered by the MCP-based electron guns [17]. Conceptually, they would allow generation of any pattern of electron current by proper exposure of the photocathode surface by either masking of the light or by several laser beams – see Fig.5. Moreover, in the latter configuration each of the EMB beamlets can be easily independently modulated – that opens additional opportunities for various applications. The first experimental attempts with easily commercially available MCPs have shown certain promise and the output electron current densities approaching 0.1 A/cm$^2$ were observed in short pulses [18], and further exploration of the MCP technologies should to be encouraged.

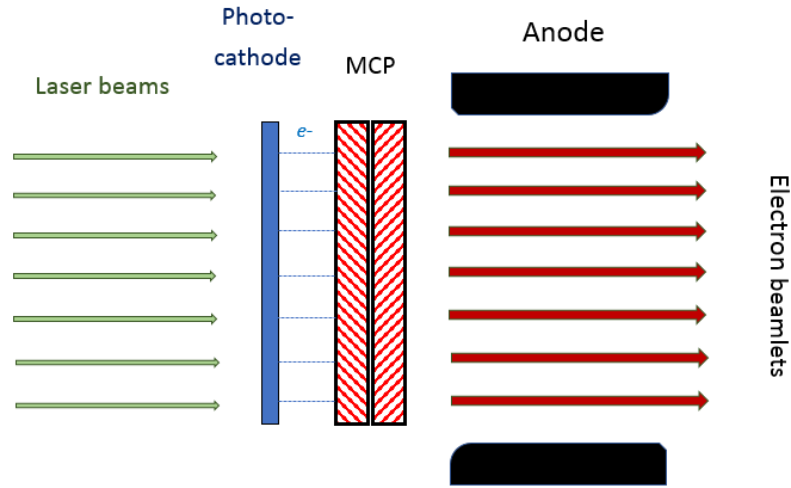

**Figure 5:** EMG generation in MCP-based electron gun.

Another effect to be taken into account is the rotation of the EMB beamlets due to their own electron space charge field Eq.(1) while propagating along the solenoidal external magnetic field $B_m$. For example, in the "electron lattice" configuration shown in Fig.2, the outer beamlets will rotate slowly (drift) around the beam central axis while staying at the same radius, i.e., the round beam remains round. The drift velocity in crossed electric and magnetic field is equal to:

$$\vec{v} = c \frac{[\vec{E}\times\vec{B}]}{B^2} \qquad (3)$$

The angle of the rotation over the distance $L$ scales as $\theta=2<j_e>L/(\beta^2_e B_m)$ and, e.g., for *average* current density of $<j_e>=1$ A/cm$^2$ it is about 4° over 0.5 m in 1T field. Of course, the axially symmetric configurations like those shown in Fig.4 will be free of such concern and will remain axially symmetric after such a drift. It is to be noted that electric field of the high-energy beam in turn also results in the orthogonal displacement of the beamlets [3, 19] and that can be used for the diagnostics of the accelerator beam internal structure.

The "electron grid" configuration can be generated out of two "electron grates" perpendicular to the high energy beam and placed one after another with the directions of the accompanying magnetic fields, and therefore, electron beamlets oriented at 90º with respect to each other.

Finally, besides generation of the low energy $O$(10 keV) electron beamlets in the electron lens type systems, the higher energy EMBs $O$(1 MeV) can also be generated out of RF photoinjectors, as it was recently demonstrated in Ref. [20].

## 4. Discussion: Applications of the Electron Multi-Beams

From the technical point of view, the electron multi-beams seem quite feasible, though their generation and practical realization may vary depending on particular application. For low energy accelerators, e.g., electron microscopes, the EM fields of the EMB $O$(1MV/m) might provide sufficiently strong impact in just one pass. In that case, the EMBs can be used in a similar fashion as glass optics elements in laser systems [21]. A notable example is microlens arrays (MLAs) which can be used for transverse optical pulse homogenization for various applications [22]. Similarly, EMBs can be used for particle beam homogenization.

For higher energy accelerators, one pass impacts of the EMBs will be small; however, in circular machines, interaction over many turns can result in significant impact. The number of turns required for substantial change of the internal distribution in high-energy beams can be estimated as $N \sim 1/\Delta Q_e$, where $\Delta Q_e$ is the betatron tuneshift due to electron beams. The electron lenses can easily induce the tune shifts of the order of $10^{-4}$-$10^{-2}$, therefore, one may expect significant effects of EMBs over hundreds to tens of thousands of turns. The speed of action and strength of impact can be significantly enhanced if the EMB current is modulated at one of the resonant (betatron or synchro-betatron) frequencies or their harmonics.

In principle, the EMBs can offer opportunities to be used for the halo control via modification of beam dynamics of larger amplitude particles; for collimation; for smearing, or homogenization of the accelerator beam phase-space; for the beam center and beam halo diagnostics as the electron beamlet positions after the interaction can be easily measured; and they may be even considered for a kind of sub-coherent beam phase space manipulations.

Some of these ideas can be tested at the electron lens setup which is being developed for the IOTA ring at Fermilab [13].

*Acknowledgements*: I would like to thank Alexey Burov and Yuri Alexahin for useful discussions on possible applications of the electron multi-beams. Fermilab is operated by Fermi Research Alliance, LLC under Contract No. DE-AC02-07CH11359 with the United States Department of Energy.